\documentclass[twocolumn,showpacs,preprintnumbers,amsmath,amssymb]{revtex4}

\usepackage{dcolumn}% Align table columns on decimal point
\usepackage{bm}% bold math
\usepackage{bbm}
\usepackage{epsfig}

\begin{document}
\preprint{APS/123-QED}

\title{Interplay between electron spin and orbital pseudospin  in  double quantum dots}

\author{Sooa Park \footnote{sooa@cqm.korea.ac.kr}}
 %\altaffiliation[Also at ]{Physics Department, Korea University.}%Lines break automatically or can be forced with \\
\author{S.-R. Eric Yang\footnote{corresponding author,  eyang@venus.korea.ac.kr}}
\affiliation{%
Dept. of Physics, Korea University
}%

\date{\today}% It is always \today, today,

\begin{abstract}
We investigate theoretically spin and orbital pseudospin  magnetic
properties  of a molecular orbital in  parabolic and elliptic
double quantum dots (DQDs). In our many body  calculation we
include intra- and  inter-dot electron-electron interactions, in
addition to the intradot exchange interaction of `p' orbitals. We
find for parabolic DQDs that, except for the half or
completely filled molecular orbital, spins in  different dots are
ferromagnetically coupled while  orbital pseudospins  are
antiferromagnetically coupled. For elliptic DQDs  spins and
pseudospins are either  ferromagnetically  or
antiferromagnetically coupled, depending on the number of
electrons in the molecular orbital. We have determined orbital
pseudospin quantum numbers for the groundstates of  elliptic DQDs.
An experiment is  suggested to test the interplay between orbital
pseudospin and spin magnetism.

\end{abstract}

\maketitle

\section{Introduction}\label{sec:introduction}

Nanometer-scale quantum dots have potential for many technological applications\cite{Petroff}.
A useful model of these systems is a parabolic quantum dot. In this model two-dimensional electrons are
confined in a parabolic potential.
The eigenenergies  of a  parabolic quantum dot  are given by
\begin{equation}\label{eq:energy:single_particle}
E_{n} = \hbar\omega(n+1), \mbox{ } n=0,\mbox{ }1, \cdots,
\end{equation}
where $\omega$ is the characteristic frequency of the parabolic
potential. This dot is rotationally invariant about the symmetry
axis of the parabolic potential. (Hereafter this axis will be
called the $z$-axis). The energy states for $n=0, 1, 2, \cdots$
are labeled as  `s', `p', `d' ... orbitals, as in the real atoms.
Each of these states is $(n+1)$-fold degenerate. The `s' orbital
is non degenerate with $z$-component of the angular momentum
quantum number $\alpha= 0$ while the `p' orbital for each dot is
doubly degenerate with $\alpha=\pm 1$.  Parabolic dots form basis
for understanding more realistic dots, for example, elliptic
dots\cite{Austing}.

When two of these dots are coupled an artificial  diatomic
molecule can be formed\cite{Wiel}. These DQDs have potential
application  in quantum computing since they may provide a basic
building block of solid state realization of  a quantum
computer\cite{Loss, EYang, Hayashi, Rushforth, Hu, Vorojtsov}.
Drawing the analogy with the transition metal
compounds\cite{Fazekas}, we expect to find interesting physics in
the  molecular orbital originating from the degenerate levels of a
quantum dot, such as the p state. The orbital degeneracy of a
shell makes the physics rich in the transition metal compounds. To
illustrate the physics we consider two sites each with a two-fold
degenerate orbital. The degenerate orbitals are denoted by $a$ and
$b$.  These  states can be labeled with a  pseudospin value: the
state $a$ with pseudospin 1/2 and the state $b$ with pseudospin
-1/2. In such a model, there are on-site repulsion energies
$U_{a}$, $U_{b}$ and $U_{ab}$, intra-atomic exchange energy $J$,
and the tunneling energies $t_a$ and $t_b$. When {\it two atoms
each contains an electron} and under the  following special case
of  $t_a =t_b =t$, $U_a =U_b =U$ and $U_{ab} =U-J$, the effective
model Hamiltonian for $t\ll U$ can be written\cite{Fazekas} as
\begin{eqnarray}
H_{eff} &=&
-\frac{4t^2}{U-2J}\left(\frac{3}{4}+\overrightarrow{S}_{1} \cdot
\overrightarrow{S}_{2} \right) \left(\frac{1}{4} -
\overrightarrow{\tau}_{1} \cdot \overrightarrow{\tau}_{2}\right)
\nonumber \\
&& - \frac{4t^2}{U}\left(\frac{1}{4}-\overrightarrow{S}_{1} \cdot
\overrightarrow{S}_{2} \right) \left(\frac{3}{4} +
\overrightarrow{\tau}_{1} \cdot \overrightarrow{\tau}_{2}\right).
\label{eq:hamiltonian:transion_metal}
\end{eqnarray}
Here,  spin and pseudospin of the electrons on the $j$-th atom are
denoted by $\overrightarrow{S}_{j}$ and
$\overrightarrow{\tau}_{j}$, respectively. This particular
Hamiltonian is rotationally invariant both in spin and pseudospin
spaces, i.e., it has the symmetry of $SU(2)\times
SU(2)$\cite{Fazekas}. In the groundstate of this Hamiltonian spins
in different atoms are {\it ferromagnetically} coupled while their
pseudospins are {\it antiferromagnetically} coupled. The
ground-state energy is $-4t^{2}/(U-2J)$.   This state has the
total spin and pseudo spin quantum numbers $S=1$ and $\tau=0$, and
is  3-fold degenerate. The first excited state with energy
$-4t^{2}/U$ has $S=0$ and $\tau=1$, and is also 3-fold degenerate.
The second excited state has zero energy with $S=\tau=0$ or
$S=\tau=1$ (10-fold degenerate).

Magnetic and excited state properties of DQDs are also of
experimental interest. Recently an experimental investigation of
orbital magnetism of DQDs containing about 50 electrons was
carried out\cite{Oosterkamp}. The single-electron tunneling
spectroscopy in a finite source drain voltage\cite{Johnson} allows
one to explore the excited states of DQDs\cite{Rontani}. This
technique is used to detect the  singlet-triplet transition in
DQDs\cite{Lee}, which plays an important role in two-qubit quantum
gates of quantum computing\cite{Loss}.

The physics of  the molecular orbital made of the  doubly
degenerate `p' states of DQDs is expected to be  more complicated
than that of the effective Hamiltonian described in
Eq.~(\ref{eq:hamiltonian:transion_metal}). In DQDs, `p' states can
be labeled with a  {\it pseudospin} value: the state with a plus
value of the $z$-component of the angular momentum has pseudospin
1/2 and the state with a minus  value has pseudospin -1/2. An
electron has thus a pseudospin 1/2 on top of ordinary spin 1/2.
Unlike transition metal compounds, since  $U_{ab}=U$ rotational
symmetry of $\vec{\tau}$ is broken, i.e., the $SU(2)\times SU(2)$
symmetry is broken in DQDs. However the system is invariant under
pseudospin rotations about the $z$-axis. It will be interesting to
investigate how  pseudospins and real spins of two dots couple in
the molecular orbital made of doubly degenerate `p' states of
DQDs. The other issue is the orbital magnetism of a
DQD\cite{Maksym, Aldea, Oosterkamp}. The orbital magnetism is
related to the $z$-component of total angular momentum.  When a
DQD contains two electrons it is zero. When the total number of
electrons in the molecular orbital $N_p$ is different from two the
$z$-component of the total angular momentum can take non-zero
values and therefore orbital magnetism may take a finite value.
For elliptic quantum dots, the rotational symmetry about the
$z$-axis is broken.

We  assume the energy level spacing of  the harmonic potential is
much larger than the strength of the Coulomb interaction:
$\hbar\omega>>U$. Only  the `p' orbitals are thus relevant (the
molecular orbital originating from the `s' orbitals are fully
occupied with four electrons and can be ignored). This can be
satisfied in self-assembled DQDs\cite{SLee} where the level
spacing is $\hbar\omega\sim 40meV$. We have obtained
wavefunctions,  energy splittings, and degeneracies of states both
for parabolic and elliptic DQDs. In our many body calculation we
include intra- and inter-dot electron-electron interactions, in
addition to the intra-dot exchange interaction of `p' orbitals.
 Since pseudospin rotational symmetry is broken
not all the eigenstates of the Hamiltonian can be also eigenstates
of the total pseudospin operator $\tau^2$.  However, we find that
the groundstates of both parabolic and elliptic DQDs happen to be
eigenstates of  $\tau^2$. We find for parabolic DQDs that, except
for the half or completely filled molecular orbital, the
groundstate spins in  different dots are ferromagnetically coupled
while  orbital pseudospins  are   antiferromagnetically coupled:
the total spin of the electrons is  maximized while  the total
pseudospin is minimized for all numbers of electrons except for
$N_p=4,8$. For elliptic DQDs  groundstate spins and pseudospins
are either  ferromagnetically or antiferromagnetically coupled,
depending on the number of electrons in the molecular orbital:
spin ferromagnetism occurs for $N_p =3$ and $N_p =5$ while
pseudospin ferromagnetism occurs for $N_p =2$ and $N_p =6$. An
experiment is  suggested to test the interplay between orbital
pseudospin and spin magnetism.

This paper is organized as follows.  In Sec. II we describe our model Hamiltonian.
In Sec.III we discuss symmetry properties of this Hamiltonian.
The proper basis states of the unperturbed Hamiltonian are given in Sec. IV.
Low-lying states of  parabolic and elliptic DQDs are given in Sec. V and VI, respectively.
Conclusions are given in Sec. VII.

\section{Model Hamiltonian for double quantum dots} \label{sec:our_model}

We describe our  model Hamiltonian for  double quantum  dots. If
the interdot exchange energy is ignored, Coulomb interactions for
the parabolic dots can be decomposed as follows:
\begin{eqnarray}
U_{\alpha,\alpha'} \!&\!=\!&\!\int\int {\rm d}^{2}x \; {\rm
d}^{2}x' |\varphi_{\alpha}({\bf x})|^{2}
|\varphi_{\alpha'}({\bf x}')|^{2}v({\bf x}-{\bf x}') \nonumber \\
V_{\alpha,\alpha'} \!&\!=\!&\!\int\int {\rm d}^{2}x \; {\rm
d}^{2}x' |\varphi_{\alpha}({\bf x})|^{2}
|\varphi_{\alpha'}({\bf x}'+{\bf d})|^{2}v({\bf x}\!-\!{\bf x}'\!-\!{\bf d}) \nonumber \\
J_{1,-1} \!&\!=\!&\!\int\int {\rm d}^{2}x \; {\rm d}^{2}x'
\varphi_{1}^{\ast}({\bf x}) \varphi_{-1}({\bf x})
\varphi_{-1}^{\ast}({\bf x}') \varphi_{1}({\bf x}') \nonumber \\
&& \times v({\bf x}-{\bf x}') \; ,
\end{eqnarray}
where $\varphi_{\alpha}$ is the wavefunction of an electron with
the $z$-component of the angular momentum $\alpha$ and $v$ is the
Coulomb potential. $U_{\alpha,\alpha'}$ is the on-dot direct
repulsion energy between two electrons, one with the $z$-component
of the angular momentum $\alpha$ and the other with $\alpha'$.
$V_{\alpha,\alpha'}$ is the same as $U_{\alpha,\alpha'}$ except
that it is the interdot direct interaction. $J_{1,-1}=J$ is the
orbital exchange energy between two electrons in different
orbitals of the same dot. Since the direct interactions are
independent of $\alpha$ and $\alpha'$, i.e.,
$U_{\alpha,\alpha'}=U$ and $V_{\alpha,\alpha'}=V$, we have the
Hamiltonian for parabolic DQDs with no hopping between dots
\begin{eqnarray}
{\cal H}_{0} &=& U\left(\sum_{j\alpha}\!n_{j\alpha 1}
n_{j\alpha,\!-\!1} +\sum_{j\sigma\sigma\prime}
n_{j1 \sigma} n_{j,\!-\!1, \sigma\prime} \!\right)\nonumber\\
 &&+ J  \!\sum_{j\sigma\sigma}c_{j1\sigma}^{\dagger}c_{j,-1,\sigma'}^{\dagger}c_{j1\sigma'}c_{j,-1,\sigma}
 \nonumber\\ &&+ V\sum_{\alpha\sigma\alpha'\sigma'}n_{1\alpha\sigma}
n_{2\alpha'\sigma'}+ E_1 \sum_{j\alpha\sigma}n_{j\alpha\sigma} \;
\label{eq:Hamiltonian_0}
\end{eqnarray}
where
$n_{j\alpha\sigma}=c_{j\alpha\sigma}^{\dagger}c_{j\alpha\sigma}$
and the fermion operator $c_{j\alpha\sigma}$
($c_{j\alpha\sigma}^{\dagger}$) annihilates (creates) the electron
with the spin $\sigma/2$ in the state of the $z$-component of the
angular momentum $\alpha$ on the $j$-th dot. The summations are
performed for $j=1,\mbox{ }2$, $\alpha=\pm 1$ and $\sigma=\pm 1$
($\uparrow$, $\downarrow$).

The hopping Hamiltonian between double dots is given by
\begin{equation}
V_{\rm hop}=-t\sum_{\alpha\sigma}
\left(c_{1\alpha\sigma}^{\dagger} c_{2\alpha\sigma} + {\rm h.c.}
\right)\; .
\end{equation}
where $t$ denotes the tunneling energy of an electron between the
opposite dots. It is assumed that the hoping of an electron
between dots changes neither the angular momentum nor the spin.
Furthermore, we assume that the hopping energy $t$ is not
dependent of the angular momentum of the electron.

The confinement potential of each  quantum dot can be assumed to
be parabolic with a distortion. The distortion an electron
experiences has an elliptic form and is represented by
\begin{eqnarray}
V_{\rm elliptic}(\vec{r})=2 \left(\frac{m^* \omega}{\hbar}\right) \epsilon xy,
\end{eqnarray}
where  $\epsilon$ is a positive parameter describing the deviation
from parabolicity. The  distortion breaks the $U(1)$ symmetry and
lifts the degeneracy of a parabolic dot. It splits the  `p'
orbitals of $E_{1}=2\hbar\omega$ with the angular momenta
$\alpha=\pm 1$ into two states
\begin{eqnarray}
|-1>'&\equiv&\frac{1}{\sqrt{2}} \left( |1> - i|-1> \right)
\; , {\rm for} \; E'_{1}=(2\hbar\omega - \epsilon) \; , \nonumber \\
|1>'&\equiv&\frac{1}{\sqrt{2}} \left( |1> + i|-1> \right)
\; , {\rm for} \; E'_{1}=(2\hbar\omega + \epsilon) \; , \nonumber \\ &&
\end{eqnarray}
where $|\alpha>$ denotes the state with the angular momentum
$\alpha=\pm 1$ and $E'_{1}$ is the single particle energy of the
elliptic dot. The  total  Hamiltonian for the DQD with the
elliptic potential can be written as
\begin{equation}
{\cal H}_{\rm tot} =  {\cal H} +  V_{\rm elliptic},
\end{equation}
where ${\cal H}={\cal H}_0+ V_{\rm hop}$ is the total Hamiltonian
for the DQD with a parabolic potential and $ V_{\rm elliptic}$
represents the Hamiltonian of the elliptic potential. The elliptic
potential term in the second quantized form is
\begin{equation}
V_{\rm elliptic} = (2\hbar\omega-\epsilon)\eta_{-1} +
(2\hbar\omega+\epsilon)\eta_{1} \; ,
\label{eq:ell1}
\end{equation}
where $\eta_{\beta}$ is the number operator of particles in the
states $|\beta =\pm 1>'$. It can be written as
\begin{eqnarray}
\eta_{\beta} &=& \frac{1}{2}\sum_{j\sigma}\left(c_{j1\sigma}-i\beta
c_{j,-1,\sigma}\right)^{\dagger} \left(c_{j1\sigma}-i\beta
c_{j,-1,\sigma}\right) \nonumber \\
&=& \frac{1}{2}\left(
\sum_{j\alpha\sigma}c_{j\alpha\sigma}^+c_{j\alpha\sigma} +
i\beta\sum_{j\alpha\sigma}\alpha
c_{j,-\alpha,\sigma}^{\dagger}c_{j,\alpha,\sigma} \right)  .
\label{eq:eta_beta}
\end{eqnarray}
In terms of pseudospin operators the elliptic potential can be
written  as
\begin{eqnarray}
V_{\rm elliptic}=2\epsilon\tau_y + E_1 N_p,
\label{eq:V_elliptic:1}
\end{eqnarray}
where
\begin{eqnarray}
\tau_y =-\frac{i}{2} \sum_{j\sigma}\left(
c_{j,1,\sigma}^{+}c_{j,-1,\sigma}
-c_{j,-1,\sigma}^{\dagger}c_{j,1,\sigma} \right) \; .
\end{eqnarray}

In general, the intradot interaction is greater than the interdot
interaction since two dots are separated with a distance $d$ each
other. One can assume that $U > J > V \gg t$. We consider elliptic
DQDs that are {\it significantly} distorted from parabolic DQDs,
so we assume that $\epsilon \gg t$. Both the elliptic potential
and the hopping terms will be treated as perturbations. The number
of `p'-orbital electrons $N_p$ that can be injected in the system
is from $0$ to $8$. For these numbers, the low energy states will
be obtained in Sec. V for parabolic dots and in Sec. VI for
elliptic dots. Before these, we will explain the symmetry
operators of our Hamiltonian and the basis states for ${\cal
H}_{0}$.

\section{Symmetry operators } \label{sec:symmetry_operators}

\underline{Orbital pseudospin operators} Let $\vec{\tau}$ be the
total orbital pseudospin operator defined by
\begin{eqnarray}
\tau^2=\frac{1}{2}(\tau_+\tau_-+\tau_-\tau_+)+\tau_z^2
\end{eqnarray}
with
\begin{eqnarray}
\tau_+&=&\tau_x+i\tau_y=\sum _{j}\tau_{j+}=
\sum_{j}\left(\sum_{\sigma}c_{j1\sigma}^+c_{j,-1,\sigma}\right)\nonumber\\
\tau_-&=&\tau_x-i\tau_y=
\sum _{j}\tau_{j-}=
\sum_{j}\left(\sum_{\sigma}c_{j,-1,\sigma}^+c_{j1\sigma}\right)\nonumber\\
\tau_z&=&\sum _{j}\tau_{jz}=\frac{1}{2}\sum_{j}\left[
\sum_{\sigma}\left(c_{j1\sigma}^+c_{j1\sigma}-c_{j,-1,\sigma}^+c_{j,-1,\sigma}\right)\right].\nonumber\\
\end{eqnarray}
These operators have the following properties.
When a  single dot contains  one electron then
$\tau_{j+}|\Downarrow>_{j}=|\Uparrow>_{j}$ and
$\tau_{j-}|\Downarrow>_{j}=0$, where $|\Uparrow>_{j}$ and
$|\Downarrow>_{j}$ denote the states of the $j$-th dot with the up
pseudospin and the down pseudospin of the electron, respectively.
If a single dot contains two electrons with opposite spins then
$\tau_{j-}|\Uparrow\Downarrow>_{j}=|\Downarrow\Downarrow>_{j}$,
where
$|\Uparrow\Downarrow>_{j}=|\Uparrow>_{j}\otimes|\Downarrow>_{j}$.
Also when two electrons with the same spin occupy a single dot then the
operation with $\tau_{j+}$ or $\tau_{j-}$ gives zero due to the
Pauli exclusion principle. Lastly in the case each dot contains one
electron we find $\tau_{+}|\Uparrow\Downarrow>=|\Uparrow\Uparrow>$
and $\tau_{-}|\Uparrow\Downarrow>=|\Downarrow\Downarrow>$, where
$|\Uparrow\Downarrow>=|\Uparrow>_{1}\otimes|\Downarrow>_{2}$.

In general, neither parabolic dot nor elliptic dot has the orbital
pseudospin number $\tau$ as a good quantum number since $[{\cal
H}_{0}, \tau^{2}]\neq 0$ holds. This can be understood as follows.
Let us define the total orbital pseudospin of $j$-th dot
$\tau_{j}$ as $\tau_{j}^2
=(\tau_{j+}\tau_{j-}+\tau_{j-}\tau_{j+})/2 + \tau_{jz}^{2}$. The
ferromagnetic exchange  term of ${\cal H}_{0}$ in
Eq.~(\ref{eq:Hamiltonian_0})  can be written as ${\cal H}_{\rm
f.ex.}=-J\frac{1}{2}[N_p - \sum_{j}(\tau_{j+}\tau_{j-} +
\tau_{j-}\tau_{j+})]$. It can be shown that this term and $\tau^2$
do not commute yielding
\begin{eqnarray}
[{\cal H}_{\rm f.ex.},\tau^{2}]&=&2J\sum_{j=1}^{2}(-1)^{j-1}[
\tau_{j+}(\tau_{1z}-\tau_{2z})\tau_{j+1,-}\nonumber \\
&& \quad -\tau_{j-}(\tau_{1z}-\tau_{2z})\tau_{j+1,+} ]\neq 0
\label{eq:commutator_J_tau}
\end{eqnarray}
where $\tau_{3,\pm} \equiv \tau_{1,\pm}$.

We will show nonetheless that for some states
\begin{eqnarray}
[{\cal H}_{\rm f.ex.},\tau^{2}]|\psi>=0. \label{eq:incomm_tau}
\end{eqnarray}
If $|\psi>$ is a one electron state of $j$-th dot then this
equation holds since ${\cal H}_{\rm f.ex.}|\psi>=0$ and
${\cal H}_{\rm f.ex.}\tau^{2}|\psi>=0$ ($\tau^{2}|\psi>$ is also a one
electron state). Also if $|\psi>$ has two electrons in a singlet
orbital-pseudospin state of the $j$-th dot
$|\tau_{j}=0,\tau_{jz}=0>=\frac{1}{\sqrt{2}}(|\Uparrow\Downarrow>_{j}-|\Downarrow\Uparrow>_{j})$
then the commutator operating to this eigenstate vanishes. It is
because the commutator in Eq.~(\ref{eq:commutator_J_tau}) has
$\tau_{j+}$, $\tau_{j-}$ or $\tau_{jz}$ in every term and those
numbers for the singlet state are all zero as follows:
$\tau_{j+}\frac{1}{\sqrt{2}}(|\Uparrow\Downarrow>_{j}-|\Downarrow\Uparrow>_{j})
=\frac{1}{\sqrt{2}}(|\Uparrow\Uparrow>_{j}-|\Uparrow\Uparrow>_{j})=0$,
$\tau_{j-}\frac{1}{\sqrt{2}}(|\Uparrow\Downarrow>_{j}-|\Downarrow\Uparrow>_{j})
=\frac{1}{\sqrt{2}}(|\Downarrow\Downarrow>_{j}-|\Downarrow\Downarrow>_{j})=0$
and
$\tau_{jz}\frac{1}{\sqrt{2}}(|\Uparrow\Downarrow>_{j}-|\Downarrow\Uparrow>_{j})=0$.
Hence, if any dot of double dots has a singlet orbital-pseudospin
states with either $\tau_{1}=0$ or $\tau_{2}=0$, the number of the
total orbital pseudospin $\tau$ becomes a good quantum number.

Meanwhile, the $z$ component of the total orbital pseudospin
commutes for parabolic dots, $[{\cal H}, \tau_{z}]=0$, so that
$\tau_{z}$ becomes a good quantum number. For elliptic dots,
however, it is not a good quantum number since $[V_{\rm elliptic},
\tau_{z}]\neq 0$.

\underline{Spin operators} Let $\vec{S}$ be the total spin
operator. Both parabolic and elliptic dots have spin rotational
symmetry, i.e., $[{\cal H}_{\rm tot},\vec{S}]=0$. This symmetry
can be verified using the spin operators
\begin{eqnarray}
S_+&=&S_x+iS_y=\sum_{j\alpha}c_{j\alpha1}^{\dagger}c_{j\alpha,-1}\nonumber\\
S_-&=&S_x-iS_y=\sum_{j\alpha}c_{j\alpha,-1}^{\dagger}c_{j\alpha1}\nonumber\\
S_z&=&\frac{1}{2}\sum_{j\alpha}(c_{j\alpha1}^{\dagger}c_{j\alpha1}-
c_{j\alpha,-1}^{\dagger}c_{j\alpha,-1}).
\end{eqnarray}

\underline{Dot pseudospin operators} One can consider another
pseudospin depending on which dot an electron belongs
to\cite{Yang, Palacios}. If the electron resides on the 1st(2nd)
dot, the dot pseudospin of it is given by
$\frac{1}{2}$($-\frac{1}{2}$). With the similar definitions to the
spin and the orbital pseudospin operators, we have
\begin{equation}
\Omega^2 = \frac{1}{2}\left(\Omega_+ \Omega_- + \Omega_- \Omega_+
\right) + \Omega_{z}^{2}
\end{equation}
with
\begin{eqnarray}
\Omega_+ &=& \Omega_x +i\Omega_y =
\sum_{\alpha\sigma}c_{1\alpha\sigma}^{\dagger}c_{2\alpha\sigma}
\nonumber \\
\Omega_- &=& \Omega_x -i\Omega_y =
\sum_{\alpha\sigma}c_{2\alpha\sigma}^{\dagger}c_{1\alpha\sigma}
\nonumber \\
\Omega_z &=&
\frac{1}{2}\sum_{\alpha\sigma}\left(c_{1\alpha\sigma}^{\dagger}c_{1\alpha\sigma}
-c_{2\alpha\sigma}^{\dagger}c_{2\alpha\sigma} \right).
\end{eqnarray}

In general, the number $\Omega$ is not a good quantum number, i.e.
$[{\cal H}_{0},\Omega^{2}]\neq0$ since the term ${\cal H}_{\rm
f.ex.}$ does not commute with $\Omega^{2}$. From the operational
calculations, however, we find the number $\Omega$ is well-defined
for some eigenstates. If $|\psi>$ is a one-electron state of the
$j$-th dot, then ${\cal H}_{\rm f. ex}|\psi>=0$. And ${\cal
H}_{\rm f. ex}\Omega^{2}|\psi>=0$ since $\Omega^{2}|\psi>$ is also
a single-electron state. Hence,
 \begin{equation}
  [{\cal H}_{\rm f.ex},\Omega^{2}]|\psi>=0 \; .
 \label{eq:incomm_Omega}
 \end{equation}
We also see that $[{\cal H}_{0}, \Omega_{z}]=0$ but $[V_{\rm hop},
\Omega_{z}] \neq 0$, and hence $\Omega_z$ does not commute with
the total Hamiltonian.

\underline{Particle-hole transformation} Both parabolic and
elliptic dots have particle-hole symmetry. For the number of
electrons in `p' orbital $N_{p}> 4$, one can straightforwardly
obtain the energy spectra and the corresponding states from the
results for the systems of $N_{p} \le 4$ via a particle-hole
transformation without detailed calculations. Here we will prove
this for the parabolic case.
%The proof for the elliptic case is similar.
Let us consider first the vacuum state in the hole space. The
maximum number of electrons that we can insert on the double dots
is 8. In this case there exists one possible state, {\it i.e.} all
orbitals are filled with particles. This fully occupied state is
defined as a hole vacuum $|0>_{h}$ which is related to the
particle vacuum $|0>$ as
\begin{equation}
|0>_h \equiv c_{111}^{\dagger}c_{11,\!-\!1}^{\dagger}
c_{1,\!-\!1,1}^{\dagger} c_{1,\!-\!1,\!-\!1}^{\dagger}
c_{211}^{\dagger} c_{21,\!-\!1}^{\dagger}
c_{2,\!-\!1,1}^{\dagger}c_{2,\!-\!1,\!-\!1}^{\dagger} |0> \; .
\end{equation}
The hole operators are defined
as $h_{j\alpha\sigma} = c_{j\alpha\sigma}^{\dagger}$
($h_{j\alpha\sigma} = c_{j\alpha\sigma}^{\dagger}$). They, of
course, fulfill the anticommutation relations
\begin{eqnarray}
&& \{ h_{j\alpha\sigma}, h_{j'\alpha'\sigma'}^{\dagger} \} =
\delta_{jj'} \delta_{\alpha\alpha'}\delta_{\sigma\sigma'} \\ && \{
h_{j\alpha\sigma}, h_{j'\alpha'\sigma'} \} = 0 \; ,
\end{eqnarray}
and operating the hole annihilation operators on $|0>_h$ gives
zero, {\it i.e.} $h_{j\alpha\sigma}|0>_h =0$. The total number of
holes is given by
\begin{equation}
N_h = \sum_{j\alpha\sigma} h_{j\alpha\sigma}^{\dagger}
h_{j\alpha\sigma} = 8-N_p \; .
\end{equation}

By replacing $c_{j\alpha\sigma}$ ($c_{j\alpha\sigma}^{\dagger}$)
with $h_{j\alpha\sigma}^{\dagger}$ ($h_{j\alpha\sigma}$) and
rearranging the operators, our Hamiltonian ${\cal H}$ can be
easily transformed as
\begin{eqnarray} \label{eq:p-h_transformation}
&&{\cal H}(U,J,V,t;N_{p})=\nonumber\\
&& {\cal H}^{h}(U,J,V,-t;N_{h})+8E_1 +(3U-J+4V)(4-N_h)\; .\nonumber\\
\end{eqnarray}
In Eq.~(\ref{eq:p-h_transformation}), the hole Hamiltonian ${\cal H}^{h}$ is defined by
\begin{equation}
{\cal H}^{h} = {\cal H}_{0}^{h} + V_{\rm hop}^{h}
\label{eq:h_hamiltonian}
\end{equation}
with
\begin{eqnarray}
{\cal H}_{0}^{h} &=& U\left(\sum_{j\alpha}\!n_{j\alpha 1}^{h}
n_{j\alpha,\!-\!1}^{h} +\sum_{j\sigma\sigma\prime}
n_{j1 \sigma}^{h} n_{j,\!-\!1, \sigma\prime}^{h} \!\right)\nonumber\\
 &&+ J  \!
 \sum_{j\sigma\sigma}h_{j1\sigma}^{\dagger}h_{j,-1,\sigma'}^{\dagger}h_{j1\sigma'}h_{j,-1,\sigma}
 \nonumber\\ &&+V \sum_{\alpha\sigma\alpha'\sigma'}n_{1\alpha\sigma}^{h}
n_{2\alpha'\sigma'}^{h}+
E_{1}^{h}\sum_{j\alpha\sigma}n_{j\alpha\sigma}^{h} \;
\end{eqnarray}
where
$n_{j\alpha\sigma}^{h}=h_{j\alpha\sigma}^{\dagger}h_{j\alpha\sigma}$
and $E_{1}^{h}=-E_{1}$, and
\begin{eqnarray}
&&V_{\rm hop}^{h} = -t\sum_{\alpha\sigma}
\left(h_{1\alpha\sigma}^{\dagger}h_{2\alpha\sigma} + {\rm h.c.}
\right)\; .
\end{eqnarray}
The energy eigenvalues for ${\cal H}^{h}$ with $N_{h}$ holes is
nothing more than the spectra for ${\cal H}$ with the same number
of electrons, which makes it possible that the low-lying energies
for $N_{p} > 4$ are obtained from the results for $N_{p} \le 4$ by
using $-t$ instead of $t$ and $N_{h}=8-N_{p}$. One can also
have the low-lying states by retransforming $h_{j\alpha\sigma}$,
$h_{j\alpha\sigma}^{\dagger}$ and $|0>_{h}$ into
$c_{j\alpha\sigma}$,  $c_{j\alpha\sigma}^{\dagger}$ and $|0>$. The
proof for the elliptic case is the same as for the parabolic one
except that
\begin{eqnarray} \label{eq:p-h_transformation2}
{\cal H}_{\rm tot}(U,J,V,t,\epsilon;N_{p})=\quad {\cal H}_{\rm tot}^{h}(U,J,V,-t,\epsilon;N_{h})\nonumber\\
+16E_1+(3U-J+4V)(4-N_h)
\end{eqnarray}
where ${\cal H}_{\rm tot}^{h}={\cal H}^{h}+V_{\rm elliptic}^{h}$
and
\begin{eqnarray}
V_{\rm elliptic}^{h}&=&-i\epsilon \sum_{j\sigma}\left(
h_{j,1,\sigma}^{\dagger}h_{j,-1,\sigma}
-h_{j,-1,\sigma}^{\dagger}h_{j,1,\sigma} \right) \nonumber \\
&&+ E_1^h N_h \; .
\label{eq:V_elliptic_h}
\end{eqnarray}
Hence, all eigenstates and energies for elliptic dots for $N_{p}
>4$ can be obtained from those for $N_{p} \leq 4$.

\section{Basis states and degenerate perturbational
calculation} \label{sec:basis_states}

\begin{figure}
\includegraphics[width=0.2\textwidth]{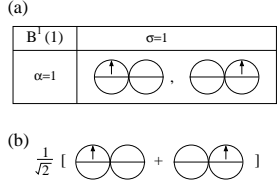}
\caption{(a)Basis set $B^{1}(N_{p}=1)$.  $\uparrow$ ($\downarrow$) is the electron 
with up (down) spin. Upper (lower) half circle
represents a state in the `p' orbital with z-component of angular
momentum $\alpha=1$($-1$). Not all the basis states are shown.
Other  states can be obtained by flipping all spins, by altering
the signs of all angular momenta, or by doing both. (b)One of the eigenstates for 
parabolic dots is shown as an example. This state has the quantum numbers $\sigma=1$ and $\alpha=1$ and
is given by a linear combination of the basis staes in
the set $B^{1}(1)$.}
\label{fig:bases:Np1}
\end{figure}

\begin{figure}
\includegraphics[width=0.3\textwidth]{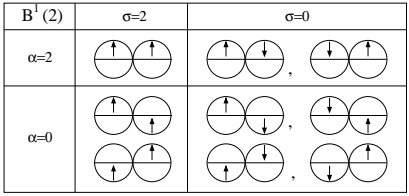}
\caption{\label{fig:10}Some basis states in the set $B^{1}(N_{p}=2)$.
States with two electrons in a single dot are not included in this basis set since we are interested only
in the low energy states.
}
\label{fig:bases:Np2}
\end{figure}

\begin{figure}
\includegraphics[width=0.4\textwidth]{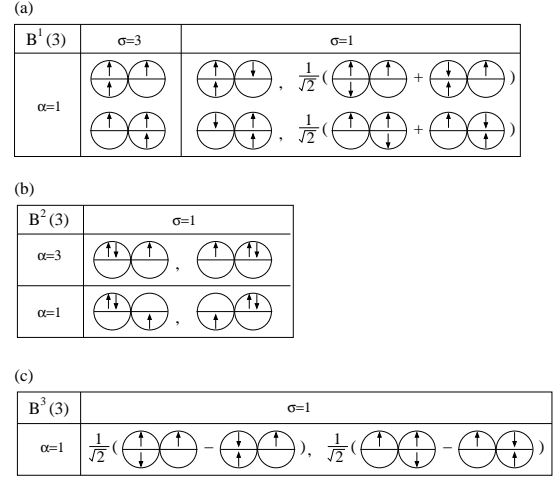}
\caption{\label{fig:9}Some  basis states in the three   sets from  $B^{1}(N_{p}=3)$ to
$B^{3}(N_{p}=3)$.}
\label{fig:bases:Np3}
\end{figure}

\begin{figure}
\includegraphics[width=0.45\textwidth]{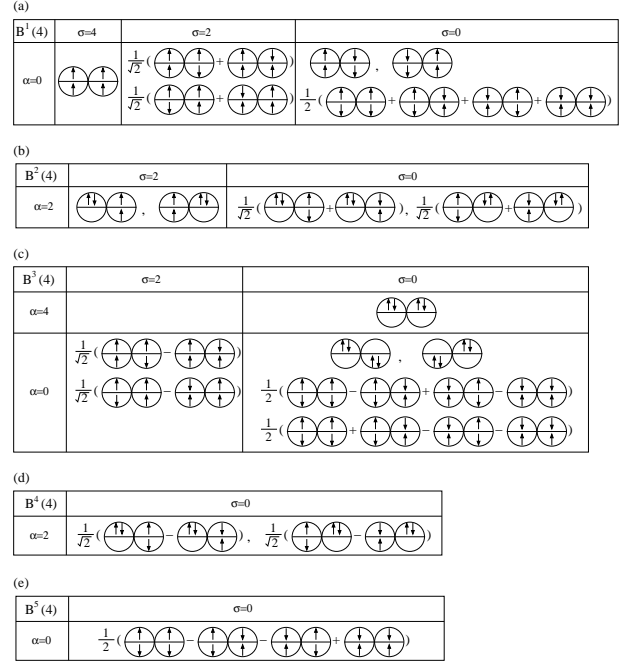}
\caption{\label{fig:9}Some  basis states in the five   sets from  $B^{1}(N_{p}=4)$ to $B^{5}(N_{p}=4)$.
}
\label{fig:bases:Np4}
\end{figure}
The total Hamiltonian of our system is ${\cal H}_0 +V_{\rm
hop}+V_{\rm elliptic}$. In the parabolic DQD $V_{\rm elliptic}$ is
absent and we will treat $V_{\rm hop}$ in degenerate perturbation
theory using the eigenstates of  ${\cal H}_0$ as basis states. In
the elliptic DQD we will use the eigenstates of  ${\cal H}_0$ to
find eigenstates of ${\cal H}_0+V_{\rm elliptic}$. Then we will
treat $V_{\rm hop}$ in degenerate perturbation theory using
eigenstates of ${\cal H}_0 +V_{\rm elliptic}$ as basis states.

In both cases we need to find the  the eigenstates of $H_0$. We
consider only the states with the energies of the lowest order of
$U$ and comment on the number of the degeneracies, $g$, of each
eigenenergy. For $N_p =1$ the eigenenergy and the degeneracy are
$(E^{(0)},g) =(0,8)$. These  basis states $B^{1}(N_{p}=1)$ are
shown in Fig.~\ref{fig:bases:Np1}. For $N_p =2$ we find
$(E^{(0)},g) =(V,16)$. These basis states $B^{1}(N_{p}=2)$ are
shown in Fig.~\ref{fig:bases:Np2}. For $N_p =3$ there are three
groups of basis states $B^{1}(N_{p}=3)$, $B^{2}(N_{p}=3)$ and
$B^{3}(N_{p}=3)$. They have $(E^{(0)},g)=(U +2V-J,24)$,
$(U+2V,16)$ and $(U + 2V+J,8)$, respectively. For $N_p =4$ there
are five groups of basis states from $B^{1}(N_{p}=4)$ to
$B^{5}(N_{p}=4)$. They have $(E^{(0)},g)=(2U+4V-2J,9)$,
$(2U+4V-J,12)$, $(2U+4V,10)$, $(2U+4V+J,4)$, and $(2U+4V+2J,1)$,
respectively.

Consider basis states $|\phi_{j}>$ that belong to  a group
$B^{l}(N_{p})$. Let us call this group $l$ and its energy
$E_{l}^{(0)}$. In the first order degenerate perturbation theory
we diagonalize the matrix
\begin{equation}
V_{i,j}^{(1)}=<\phi_{i}| V_{\rm pert} |\phi_{j}>,
\label{eq:V_matrix:1st}
\end{equation}
and find the eigenstates and energies. If this matrix is zero we
apply the 2nd order degenerate perturbation theory by
diagonalizing the matrix
\begin{equation}
V_{i,j}^{(2)}=<\phi_{i}| V_{\rm pert} \frac{1}{E_{l}^{(0)}-{\cal
H}_{\rm unpert}} V_{\rm pert} |\phi_{j}> \; .
\label{eq:V_matrix:2nd}
\end{equation}
Note that the state $V_{pert} |\phi_{j}>$ may not belong to the
group $l$; it may belong to another group $l'$ with degenerate
energy $E_{l'}$. The resulting eigenvectors are the linear
combinations of the basis states $|\phi_{j}>$ with appropriate
expansion coefficients $c_j$:
$|\psi_{k}>=\sum_{j}c_{j}|\phi_{j}>$. The perturbed eigenenergies
are $<\psi_{k}| V_{\rm pert} \frac{1}{E^{l}-{\cal H}_{\rm unpert}}
V_{\rm pert} |\psi_{k}>$. In the parabolic DQDs when the number of
electrons is odd, the degeneracy of the low energy states for
${\cal H}_{0}$ is lifted in the 1st order perturbation of $V_{\rm
hop}$. For a even number of electrons, on the other hand, we need
to use the 2nd order energy. It indicates that the hoppings of
electrons are more important for the odd-numbered system than for
the even-numbered one.  All the matrix elements $V_{ij}$  have
common proportionality factor $t$ or $t^2$, and consequently the
expansion coefficients $c_i$ in the eigenvectors are all
independent of these factors. For an elliptic potential degenerate
perturbation theory is applied twice. First we apply it to $
V_{\rm elliptic} $, and then apply it to $V_{\rm hop} $ using the
eigenstates of ${\cal H}_{0}+V_{\rm elliptic}$.

We have {\it tested} our degenerate perturbation theory for the
following special case of  $t_a =t_b =t$, $U_a =U_b =U$ and
$U_{ab} =U-J$.  We have calculated the eigenstates, eigenenergies,
degeneracies, and quantum numbers. These results are consistent
with those for the Hamiltonian of the transition metal oxides in
Eq.~(\ref{eq:hamiltonian:transion_metal}).

\section{Low-lying energy states for parabolic quantum dots}\label{sec:parabolic_dots}

The parabolic DQD is described by the Hamiltonian $H_0+V_{hop}$.
Here the low lying energies and the eigenstates of our DQD model
are investigated by using the degenerate perturbation theory for
$N_p \le 4$ and by transforming electrons to holes for $N_p \ge
5$. From now on, in order to avoid the long expression, we drop
the constant term $E_1 N_p$ in the Hamiltonian ${\cal H}_{0}$ when
writing the energy eigenvalues $E^{(0)}$ for ${\cal H}_0$ and $E$
for ${\cal H}$. We denote the $N_{p}$-particle eigenfunction for
the parabolic Hamiltonian ${\cal H}$ as
$|\psi_{\alpha_{T},\sigma_{T}}^{(k)}(N_{p})>$ where $\alpha_{T}$
and $\sigma_{T}$ indicate the $z$-component of the total angular
momentum and the twice of the total spin, respectively. In order
to distinguish the states with the same values of $\alpha_{T}$ and
$\sigma_{T}$, we do numbering by assigning $k$. The state with the
lower $k$ value has either the same or the lower energy than the
higher $k$-valued one. Since the hopping term can be written as
$V_{\rm hop} = -2t \Omega_{x}$ we  need to diagonalize the first
order matrix elements  $(\Omega_x)_{i,j}^{(1)}$. If these first
order matrix elements are all zero we need to diagonalize the
second order matrix elements $(\Omega_x)_{i,j}^{(2)}$ (the
definitions of the superscript 1 and 2 are given in
Eqs.~(\ref{eq:V_matrix:1st}) and (\ref{eq:V_matrix:2nd})).

\begin{figure}
\includegraphics[width=0.45\textwidth]{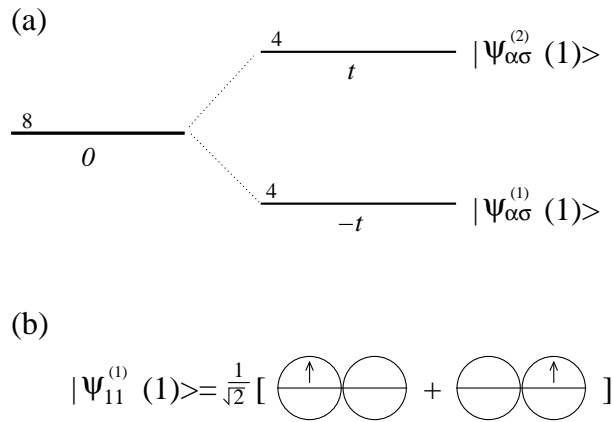}
\caption{The low-lying energy states for $N_p =1$ are shown.  Left
and right horizontal bars indicate energy levels of ${\cal H}_0$
and ${\cal H}_0+V_{\rm hop}$.  The numbers on the bars indicate
the number of degeneracy.  (a)The energy spectrum and the states
are shown. Using the particle-hole transformation, we can obtain
the energy spectrum  for $N_p =7$ from that of $N_p =1$, see the
text. (b) One of the groundstates is depicted.}
\label{fig:parabolic:1}
\end{figure}

\underline{$N_p =1$}: The energy spectrum is depicted in
Fig.~\ref{fig:parabolic:1} (a). Since $[V_{\rm hop},
\Omega_{z}]\neq 0$ leads to the broken dot pseudospin rotational
symmetry, the energy is split into two levels. Explicit
calculations show that all eigenstates have $\tau=1/2$, $S=1/2$
and $\Omega=1/2$. Each split level is 4-fold degenerate with
$\tau_{z}=\pm 1/2$ ($\alpha=\pm 1$) and $S_{z}=\pm 1/2$
($\sigma=\pm 1$). For the ground states
$|\psi_{\alpha\sigma}^{(1)}>$ we have $\Omega_{x}=1/2$. One of the
ground states, which has $\tau_{z}=1/2$ and $S_{z}=1/2$, is shown,
as an example, in Fig.~\ref{fig:parabolic:1} (b). Other ground
states can be obtained by flipping the orbital-pseudospin or the
spin, or doing both in the example state. For the excited states
$|\psi_{\alpha\sigma}^{(2)}>$, we have $\Omega_{x}=-1/2$. For all
these eigenstates $[{\cal H}_{0}, \tau^{2}]
|\psi_{\alpha_{T},\sigma_{T}}^{(k)}(N_{p})>=
 0$ although
the operational identity is as $[{\cal H}_{0},\tau^{2}]\neq 0$.
This is because  the eigenstates are one electron states. See
Eq.~(\ref{eq:incomm_tau}) and the explanation following it.
Consequently, $\tau$ is a  good quantum number. For the same
reason $\Omega$ is a good quantum number (see
Eq.~(\ref{eq:incomm_Omega})). The spectrum for $N_p =7$ is exactly
the same as that of $N_p =1$  except that the eigenstates
corresponding to $|\psi_{\alpha\sigma}^{(1)}>$ have the energy
$E=9U-3J+12V+t$ while $|\psi_{\alpha\sigma}^{(2)}>$ have
$E=9U-3J+12V-t$. This is because   $t$  is replaced by $-t$  in
the particle-hole transformation.

\begin{figure}
\includegraphics[width=0.45\textwidth]{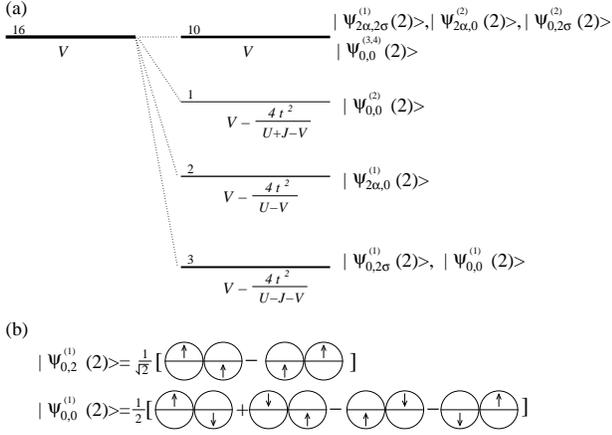}
\caption{The low-lying energy states for $N_p =2$ are shown: (a) the energy
spectrum and the states, and (b) two of the ground states are depicted.}
\label{fig:parabolic:2}
\end{figure}

\underline{$N_p =2$}: In the basis states  we include only states
with one electron in each dot. With these basis states  the
Hamiltonian matrix of  $V_{\rm hop}$ vanishes since the hopping
term produces states with two electrons on the same dot. Therefore
we apply  the 2nd order degenerate perturbational calculation to
find eigenstates.  The results are shown  in
Fig.~\ref{fig:parabolic:2} (a). For the ground states, we find
$S=1$ and $\Omega=1$. It is 3-fold degenerate with $S_z =\pm 1$
and $S_z =0$. Fig.~\ref{fig:parabolic:2}(b) shows the graphical
representations for the ground states with $S_z =1$
($\sigma_{T}=2$) and $S_z =0$ ($\sigma_{T}=0$). The states with
$S_z =-1$ can be obtained by flipping all spins of the state with
$S_{z}=1$. Note that for all eigenstates each dot contains only
one electron. Therefore from Eqs.~(\ref{eq:incomm_tau}) and
~(\ref{eq:incomm_Omega}) it follows that the numbers $\tau$ and
$\Omega $ are good quantum numbers of the eigenstates.  There are
competitions between ferromagnetic and superexchange interactions
in our model Hamiltonian. Two-electron parabolic DQDs are stable
when the ferromagnetic interactions win, making the spins of
electrons parallel. On the other hand, the orbital pseudospins are
antiparallel so that $\tau=0$.

\begin{figure}
\includegraphics[width=0.45\textwidth]{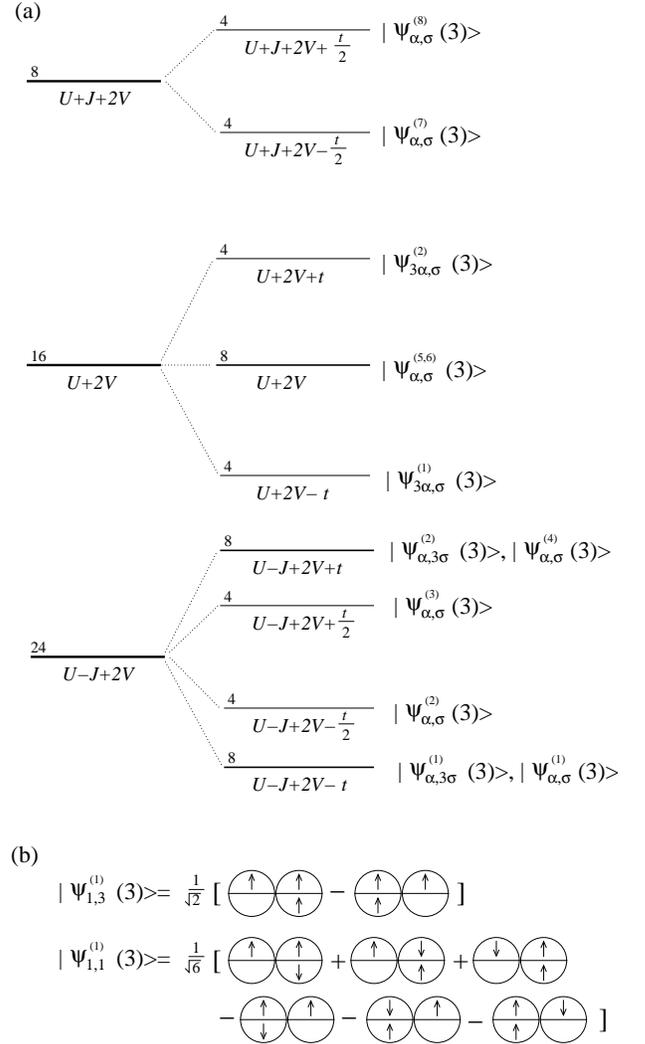}
\caption{The low-lying energy states for $N_p =3$ are shown: (a) the energy
spectrum and the states, and (b) two of the ground states are depicted.}
\label{fig:parabolic:3}
\end{figure}

\underline{$N_p =3$}: Fig.~\ref{fig:parabolic:3} (a) shows how the
low energy spectrum for $N_p =3$ is changed by the perturbation
$V_{\rm hop}$.  Note that in the groundstates each dot contains
either one or two electrons and the dot with two electrons has the
orbital pseudospin singlet state of $\tau_{j}=0$. Therefore from
Eq.~(\ref{eq:incomm_tau}) it follows that $\tau $ is a good
quantum number of the groundstates. $\Omega$ also happens to be a
good quantum number for the groundstates. The ground-state energy
corresponds to the states with $\tau=\frac{1}{2}$, $S=\frac{3}{2}$
and $\Omega=\frac{1}{2}$. They have $S_z =\pm\frac{1}{2}$ or $S_z
=\pm\frac{3}{2}$ and $\tau_z =\pm\frac{1}{2}$, respectively,
yielding the ground state 8-fold degenerate. In
Fig.~\ref{fig:parabolic:3} (b), we show two ground states of them,
one with $\tau_z =\frac{1}{2}$ and $S_z =\frac{3}{2}$ and the
other with $\tau_z =\frac{1}{2}$ and $S_z =\frac{1}{2}$. The
others can be obtained by flipping all spins or all orbital
pseudospins or both. From the ground states, one can notice that
our three-electron system favors the total spin of electrons to be
maximized, showing the spin ferromagnetism. On the other hand, the
orbital and dot pseudospins are minimized. Not all the eigenenergy
states are eigenstates of $\Omega^2$, for example,
$|\psi_{\alpha,\sigma}^{(2,3)}>$. But the numbers $S$, $S_z$ and
$\tau_z$ are good quantum numbers of all the eigenenergy states.

\begin{figure}
\includegraphics[width=0.45\textwidth]{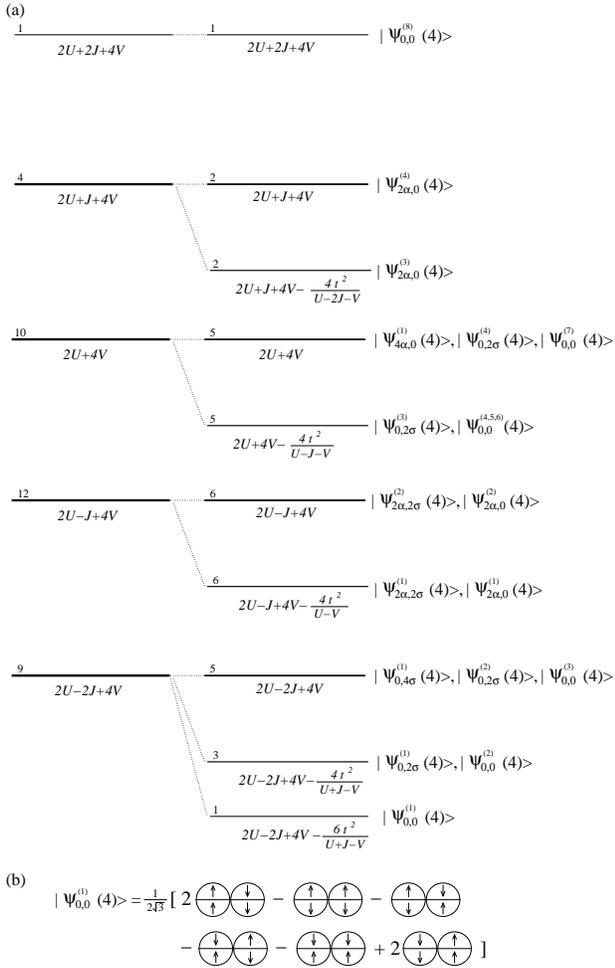}
\caption{The low-lying energy states for $N_p =4$ are shown: (a) the energy
spectrum and the states, and (b) the ground state is depicted.}
\label{fig:parabolic:4}
\end{figure}

\underline{$N_p =4$}: After the 2nd-order perturbation
calculations, we obtain the energy spectra in
Fig.~\ref{fig:parabolic:4} (a). The bases $B^{1}(4)$ with the
lowest unperturbed energy $E^{(0)}=2U-2J+4V$ are recombined to be
the eigenstates with the ground-state energy
$E=2U-2J+4V-6t^{2}/(U+J-V)$, and the others with
$E=2U-2J+4V-4t^{2}/(U+J-V)$ and $E=2U-2J+4V$. The ground state
$|\psi_{0,0}^{(1)}(4)>$ is unique and has $\tau_z =0$, $S=0$ and
$S_z =0$. It is also the eigenstate for $\tau$ with $\tau=0$.
 Note that in the
groundstates each dot contains  two electrons and both dot have
the singlet states of the orbital pseudospins
$\tau_{1}=\tau_{2}=0$. Therefore from Eq.~(\ref{eq:incomm_tau}) it
follows that $\tau $ is a good quantum number of the groundstates.
These results suggest that the system with $N_p =4$ is stable when
the spins and the orbital pseudospins of electrons are
antiparallel each other. The reason is as follows. For $N_p =4$,
the electrons occupy four different quantum states of double dots.
Unlike the case for $N_p =1$, $2$ and $3$, whenever an electron of
one dot hops onto the other dot it must face another electron with
the opposite spins due to the Pauli exclusion principle. Hence,
the superexchange interaction dominates the ferromagnetic
interaction in order to lower energy by hopping. It is worth
noting that, in the ground state, $S_1 =S_2 =1$. Within a dot, the
spins are ferromagnetic. However, $\overrightarrow{S}_1$ and
$\overrightarrow{S}_2$ are aligned in opposite directions, leading
$S=0$. Furthermore, the total orbital pseudospin $\tau$ is
naturally zero. In the excited states, $\tau_z$, $S$ and $S_z$ are
good quantum numbers.

\underline{$5\le N_p \le 8$}: The spectra for $N_p =5$, $6$ and
$7$ are obtained from those for $N_p=3$, $2$ and $1$,
respectively, by the particle-hole transformation as we mentioned
in Sec.~\ref{sec:symmetry_operators}. The procedure is as below.
First, we obtain the energy spectra of ${\cal
H}^{h}(U,J,V,t;N_{h}=8-N_{p})$, which is nothing but the spectra
of ${\cal H}(U,J,V,t;N_{p}=N_{h})$. And each eigenstate of ${\cal
H}^{h}$ is given by simply writing $h_{j\alpha\sigma}$
($h_{j\alpha\sigma}^{\dagger}$) instead of $c_{j\alpha\sigma}$
($c_{j\alpha\sigma}^{\dagger}$) and $|0>_{h}$ instead of $|0>$ in
the eigenstate of ${\cal H}$. Then, we change the hopping energy
from $t$ to $-t$ in order to obtain the energy spectrum for ${\cal
H}^{h}(U,J,V,-t;N_{h})$. Finally we add $(3U-J+4V)(4-N_h )$ to the
energy and transform all states by replacing $h_{j\alpha\sigma}$
($h_{j\alpha\sigma}^{\dagger}$) with $c_{j\alpha\sigma}^{\dagger}$
($c_{j\alpha\sigma}$) and $|0>_h$ with $c_{111}^{\dagger}
c_{11,-1}^{\dagger} c_{1,-1,1}^{\dagger} c_{1,-1,-1}^{\dagger}
c_{211}^{\dagger} c_{21,-1}^{\dagger} c_{2,-1,1}^{\dagger}
c_{2,-1,-1}^{\dagger}|0>$. For $N_p =8$, all of the quantum states
on the double dots are filled up and the energy of the system is
$E=12U-4J+16V$. The ground states for $N_p >4$ have the same
magnetic properties for the states with $8-N_p$. This results from
the fact that $(2N_p -8)$ particles among $N_p$ electrons always
remain nonmagnetic, each two bound as a pair of electrons with up
and down spins in the same orbitals on the same dots. For $N_p
=8$, the system is naturally nonmagnetic.

\section{Low-lying energy states of elliptic Potential}\label{sec:elliptic_dots}

In the following  we will investigate the energy spectra for
elliptic dots from $N_{p}=1$ to $N_{p}=4$. We obtain the total
energies and the eigenstates up to the nonvanishing leading terms
using two successive degenerate perturbation calculations, the
first, with respect to $V_{\rm elliptic}$ and the second, to
$V_{\rm hop}$, as we mentioned in Sec. IV. In these calculations,
the term $E_{1} N_{p}$ in the potential $V_{\rm elliptic}$ as well
as in the Hamiltonian ${\cal H}_{0}$ are ignored.

We will denote the final eigenstates for ${\cal H}_{\rm tot}$ as
$|\psi_{\sigma_{T}}^{\prime(k)}(N_{p})>$ where $\sigma_{T}$ and
$k$ are equivalently defined as in Sec.~\ref{sec:parabolic_dots}.
Note that $\alpha_{T}$ is dropped this time because $V_{\rm
elliptic}$ breaks the orbital pseudospin's up-down symmetry so
that the $z$-component of the total angular momentum is not
conserved anymore.

In the first order perturbational calculation with respect to
$V_{\rm elliptic}$ the matrix $(V_{\rm
elliptic})_{i,j}^{(1)}\propto (\tau_y)_{i,j}^{(1)}$ has to be
diagonalized (see Eq.~(\ref{eq:V_matrix:1st})). When all matrix
elements of $(V_{\rm elliptic})_{i,j}^{(1)}$ vanish, the second
order matrix elements $(V_{\rm elliptic})_{i,j}^{(2)}$ is
calculated (see Eq.~(\ref{eq:V_matrix:2nd})).

\begin{figure}
\includegraphics[width=0.45\textwidth]{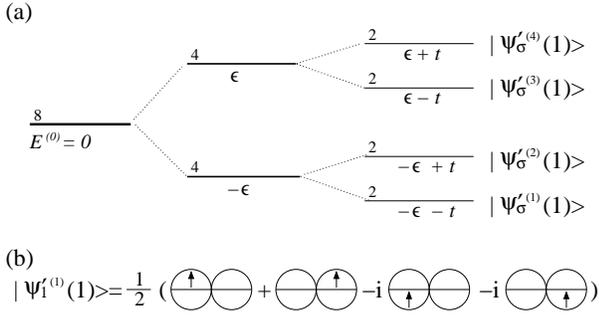}
\caption{(a)The low-lying energy spectrum for $N_p =1$ is depicted
for an elliptic DQD. The left, middle, and right horizontal bars
indicate energy levels of ${\cal H}_0$, ${\cal H}_0+V_{\rm
elliptic}$, and ${\cal H}_{\rm tot}={\cal H}_0+V_{\rm
elliptic}+V_{\rm hop}$. (b)One of the groundstates for ${\cal
H}_{\rm tot}$ is depicted. } \label{fig:elliptic:Np1}
\end{figure}

\underline{$N_p =1$}: The resulting spectra for $N_p =1$ are shown
in Fig.~\ref{fig:elliptic:Np1} (a). For $N_{p}=1$, the Hamiltonian
${\cal H}_{\rm tot}$ is exactly diagonalized. Unlike the
Hamiltonian for parabolic dots ${\cal H}_{0}$, the total
Hamiltonian does not commute with the $z$-component of the total
orbital pseudospin, i.e., $[{\cal H}_{\rm tot},\tau_{z}]\neq 0$.
This is due to the elliptic potential $V_{\rm elliptic}$. Hence,
after the perturbation $V_{\rm elliptic}$ is introduced, the
energy is split into two levels. Each level is split again into
two levels by the hopping term $V_{\rm hop}$, which is responsible
for the fact that ${\cal H}_{\rm tot}$ does not commute with
$\Omega_{z}$, i.e., $[{\cal H}_{\rm tot},\Omega_{z}]\neq 0$ as we
mentioned in the previous section~\ref{sec:parabolic_dots}. The
ground state is depicted in Fig.~\ref{fig:elliptic:Np1} (b). The
other ground state can be obtained by flipping the electron's
spin. For these ground states $|\psi_{\sigma}^{\prime(1)}>$, we
have $(\tau,\tau_y)=(\frac{1}{2},-\frac{1}{2})$, $(S,S_z
)=(\frac{1}{2},\frac{1}{2}\sigma)$ and $(\Omega,\Omega_x
)=(\frac{1}{2},\frac{1}{2})$. Note that in all eigenstates each
dot contains either one or zero electrons. Therefore from
Eqs.~(\ref{eq:incomm_tau}) and (\ref{eq:incomm_Omega}) it follows
that $\tau $ and $\Omega $ are good quantum numbers. For the
excited states, all eigenstates have $\tau =\frac{1}{2}$,
$S=\frac{1}{2}$ and $\Omega=\frac{1}{2}$. Each energy is doubly
degenerate with $S_z =\pm \frac{1}{2}$.

\begin{figure}
\includegraphics[width=0.45\textwidth]{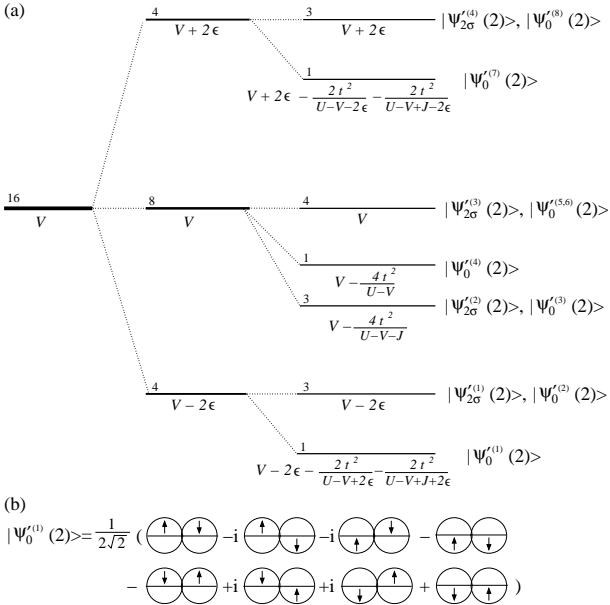}
\caption{(a)The low-lying energy spectrum for $N_p =2$ of an
elliptic DQD. (b)The groundstate is depicted. }
\label{fig:elliptic:Np2}
\end{figure}

\underline{$N_p =2$}: The energy spectra of ${\cal H}_{\rm tot}$
for $N_p =2$ are shown in Fig.~\ref{fig:elliptic:Np2} (a). Note
that similar to the case of  two-electron parabolic DQDs, the
numbers $\tau$ and $\Omega$ are good quantum numbers. The ground
state $|\psi^{\prime(1)}_{0}(2)>$ is unique and is drawn in
Fig.~\ref{fig:elliptic:Np2} (b). It has $(\tau,\tau_y) =(1,-1)$,
$(S=0,S_z) =(0,0)$ and $\Omega=1$. One can notice that the
eigenvalues of the total orbital pseudospin and the total spin are
different from those of the ground state of a parabolic DQD with
$\tau=0$ and $S=1$. Excited states
$|\psi^{\prime(2)}_{2\sigma}(2)>$ and $|\psi^{\prime(3)}_{0}(2)>$
of elliptic DQDs have the same form as the degenerate groundstates
$|\psi_{0,2\sigma}^{(1)}(2)> $ and $|\psi_{0,0}^{(1)}(2)> $ of
parabolic DQDs. In these ground states of parabolic DQDs, the
total orbital pseudospin of two electrons forms a singlet state,
$(|\Uparrow\Downarrow>-|\Downarrow\Uparrow>)/\sqrt{2}$  with
$\tau=0$. Since $V_{\rm elliptic}$ flips all pseudospins in turn,
this singlet state vanishes as
$2\epsilon\tau_{y}\frac{1}{\sqrt{2}}(|\Uparrow\Downarrow>-|\Downarrow\Uparrow>)
=-i\epsilon\frac{1}{\sqrt{2}}(|\Downarrow\Downarrow>+|\Uparrow\Uparrow>-
|\Uparrow\Uparrow>-|\Downarrow\Downarrow>)=0$. Hence, the
antisymmetric property of the orbital pseudospin singlet leads to
the unchanged  total energy. In the groundstate of a  two-electron
elliptic DQD, the total spin is minimized while the total orbital
pseudospin is maximized, showing orbital pseudospin
ferromagnetism.

\begin{figure}
\includegraphics[width=0.45\textwidth]{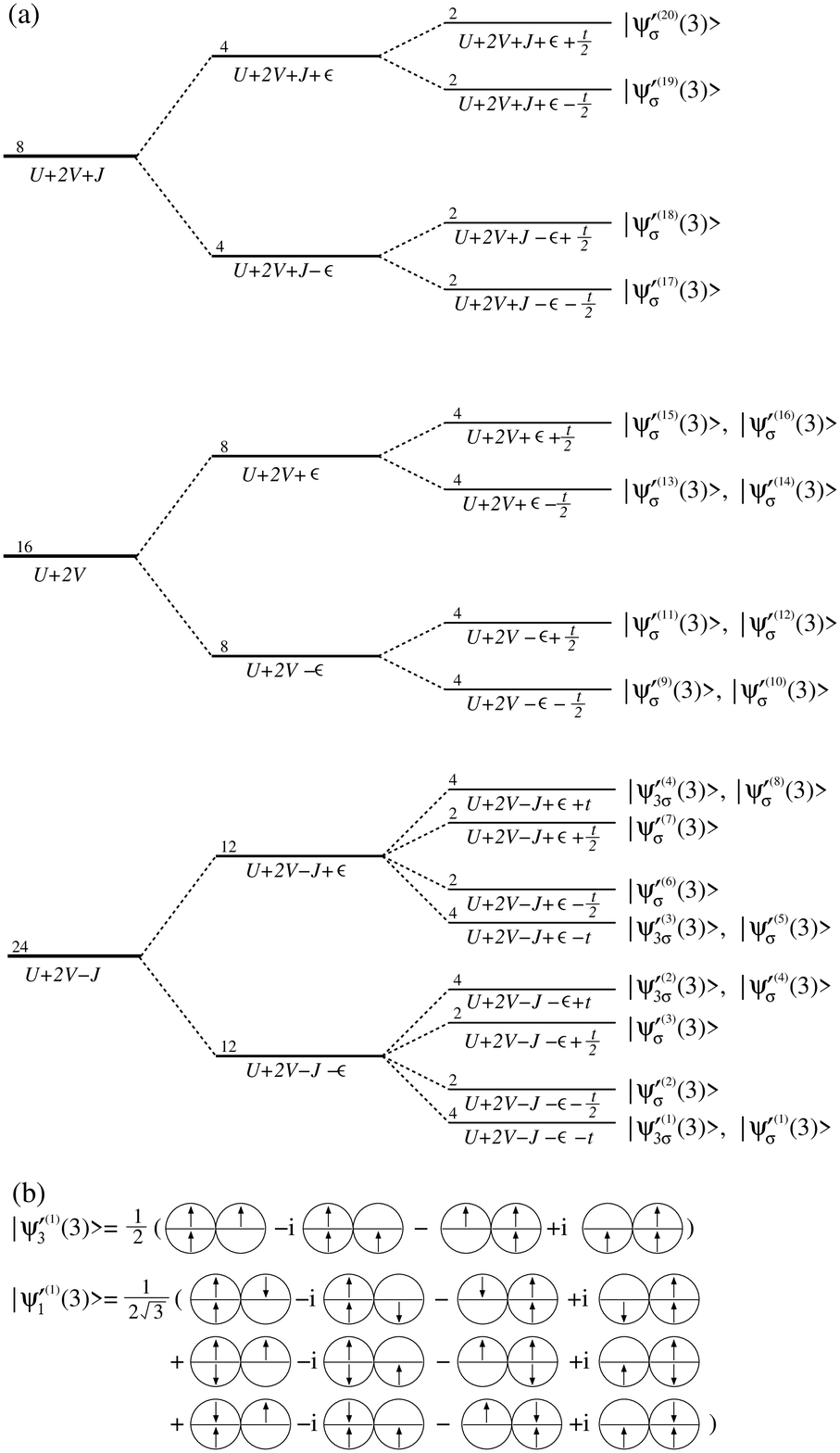}
\caption{(a)The low-lying energy spectrum for $N_p =3$ of an
elliptic DQD. (b)Two ground states are depicted. }
\label{fig:elliptic:Np3}
\end{figure}

\underline{$N_p =3$}: The results of the elliptic DQD for $N_p =3$
are shown in Fig.~\ref{fig:elliptic:Np3} (a). Each unperturbed
energy is split into two levels by $V_{\rm elliptic}$: the lower
energy state has $\tau_y =- \frac{1}{2}$ and the higher energy
state   $\tau_y =+ \frac{1}{2}$. They are split again when  the
hopping perturbation $V_{\rm hop}$ is introduced, as shown in the
spectra in the right side. The ground states are four-fold
degenerate with $S_z =\pm\frac{1}{2}$ and $S_z =\pm\frac{3}{2}$.
Two of them are depicted in Fig.~\ref{fig:elliptic:Np3} (b), and
others can be obtained by flipping all spins in the depicted
states. In the ground states, we have $(\tau,\tau_y)
=(\frac{1}{2},-\frac{1}{2})$, $S=\frac{3}{2}$ and
$(\Omega,\Omega_x)=(\frac{1}{2},\frac{1}{2})$. For the same reason
as in the parabolic DQD for $N_p =3$, $\tau$ and $\Omega$ are good
quantum numbers in the ground states. Similar to the parabolic
DQD, the ground states show the spin ferromagnetism while the
orbital and the dot pseudospins are minimized. This can be
understood as follows: the ground states for elliptic dots are the
linear combinations of the parabolic-dot ground states with the
same spin quantum number $\sigma_T$, i.e.,
$|\psi^{\prime(1)}_{3\sigma}(3)>=\frac{1}{\sqrt{2}}(|\psi^{(1)}_{1,3\sigma}(3)>-i|\psi^{(1)}_{-1,3\sigma}(3)>)$
and
$|\psi^{\prime(1)}_{\sigma}(3)>=\frac{1}{\sqrt{2}}(|\psi^{(1)}_{1,\sigma}(3)>-i|\psi^{(1)}_{-1,\sigma}(3)>)$.
Hence, it is natural that they show the same magnetic properties
of the ground states of parabolic dots. Note that the numbers $S$
and $S_z$ are good quantum numbers in the excited states.

\begin{figure}
\includegraphics[width=0.45\textwidth]{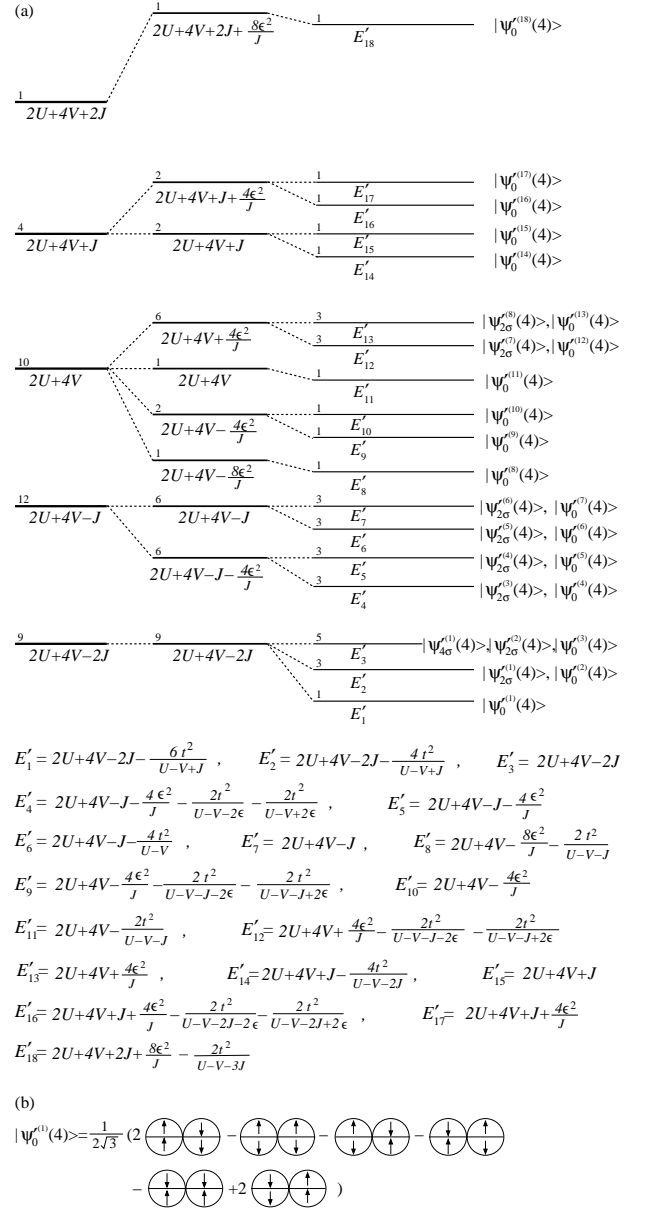}
\caption{(a)The low-lying energy spectrum for $N_p =4$ of an
elliptic DQD. (b)The ground state is depicted. }
\label{fig:elliptic:Np4}
\end{figure}

\underline{$N_p =4$}: Fig.~\ref{fig:elliptic:Np4} (a) shows the
energy spectra for $N_p =4$. The energy of $E^{(0)}=2U+4V-2J$ and
the corresponding basis states $B^{1}(4)$ are not changed after
the perturbation $V_{\rm elliptic}$. The reason is as follows.
Each dot in the basis states has $S_j =1$, $j=1,2$ (triplet
states). The spins of two electrons within a dot are parallel
which does not permit the change of the orbital pseudospins of the
electrons due to Pauli exclusion principle. Hence, the elliptic
potential does not change the basis states $B^{1}(4)$. Since only
the hopping term $V_{\rm hop}$ has an effect on those states, the
ground state is the same as that for the parabolic dots. We depict
it in Fig.~\ref{fig:elliptic:Np4} (b). Of course, $S=0$ and
$\tau=0$ in the ground state. In the excited states, $S$ and $S_z$
are, in general, good quantum numbers.

\underline{$5\le N_p \le 8$}: The elliptic potential is unchanged
under the particle-hole transformation.
The energy spectra and the
states for $N_p \geq 5$ can thus be obtained from those for $N_p \leq
4$ via the particle-hole transformation.

\section{Conclusion and discussion}\label{sec:conclusion}

\begin{center}
\begin{table}
\begin{tabular}{|c|c|c|c|c|c|c|c|c|} \hline
 $N_p$  & 1 & 2 & 3  & 4 & 5 & 6 &   7 &8 \\  \hline
 $S$   & 1/2 &1 &  3/2 & 0  &  3/2  &   1  &    1/2 &0  \\  \hline
  $\tau$ & 1/2  &   0 &  1/2 &   0&    1/2 &   0 & 1/2 &0   \\ \hline
 $\Omega$   & 1/2 &   1 & 1/2 &   - &  1/2 &   1 & 1/2 &0   \\  \hline
  $\Omega_x$ & 1/2 & -   &   1/2  & - & 1/2 & - & 1/2 & 0 \\ \hline
\end{tabular}
\caption{Groundstate magnetic properties of parabolic DQDs for
different values of $N_p$. A hyphen means that the relevant
quantum number is not a good quantum number.}
\end{table}
\end{center}

\begin{center}
\begin{table}
\begin{tabular}{|c|c|c|c|c|c|c|c|c|} \hline
 $N_p$  & 1  & 2  &  3  & 4  & 5   &  6 &   7 &8 \\  \hline
 $S$    & 1/2& 0  & 3/2 & 0  & 3/2 &  0 &  1/2  &0  \\  \hline
 $\tau$ & 1/2& 1  & 1/2 &   0& 1/2 &  1 &    1/2  &0  \\ \hline
 $\tau_{y}$ & -1/2 & -1 &  -1/2 & 0 & -1/2 & -1 & -1/2 & 0\\ \hline
 $\Omega$   & 1/2 &  1 &    1/2 &  - & 1/2& 1 & 1/2 &0  \\  \hline
 $\Omega_x$ & 1/2 & -  & 1/2 & - & 1/2 & - & 1/2 & 0 \\ \hline
\end{tabular}
\caption{Groundstate magnetic properties of elliptic  DQDs for
different values of $N_p$.}
\end{table}
\end{center}
We have investigated theoretically low excitation  states in the
`p' molecular orbital of parabolic and elliptic DQDs.
In our many body  calculation we include intra- and  inter-dot
electron-electron interactions, in addition to the intradot
exchange interaction of `p' orbitals. Wavefunctions,  energy
splittings, and degeneracies of states  of  the molecular orbitals
are determined.
The low lying excited states may be probed experimentally by measuring
current when a finite source drain voltage is present\cite{Johnson,Lee}.
Our result shows how many current peaks should be observed in such an experiment.

We have also  determined  the
properties of orbital  magnetism of groundstates.
Since the magnetization is directly related to the z-component of
the total angular momentum  our results may be used to determine
the value of the magnetization of DQDs\cite{Oosterkamp}.  Note that the z-component of the total
angular momentum is in fact equal to the value of
the z-component of the total pseudospin.
In elliptic DQDs  the z-component of the total angular momentum is
not a good quantum number.

The orbital pseudospin and spin properties  for parabolic and
elliptic DQDs are summarized in Tables I and II, respectively. We
see for parabolic DQDs that the total spin of electrons takes the
maximum possible value  except for  $N_p =4$ and $N_p =8$. For
these values of  $N_p $  the total spin is minimum. The orbital
pseudospin is zero for $N_p =2,4,6,8$. Note that  for both $N_p=2$
and $6$ {\it the spins and pseudospins are ferromagnetic and
antiferromagnetic, respectively}, i.e., $S=1$ and $\tau=0$.
However, for elliptic DQDs the opposite is true. In this case the
groundstate energy is minimized when the total spin takes the
value $S=0$. This difference in the interplay between orbital
pseudospin and spin magnetism in parabolic and elliptic DQDs can
be tested experimentally. When a weak magnetic field is applied
the groundstate energy of the parabolic DQD should split into
three while that of the elliptic dot remains unsplit. A sufficient
deformation from the parabolicity is required since the strength
of the ellipticity assumed to be stronger than the magnitude of
the tunneling energy in our results. In our calculations we have
assumed that the exchange interaction between dots is negligible.
In some cases the interdot exchange interaction can be important,
and can lead to canted phases\cite{Sanchez}. Further
investigations including the interdot exchange are needed to
test the stability of  canted states  in our DQDs.

This work is supported by the Korea Science and Engineering
Foundation (KOSEF)  through the Quantum Functional Semiconductor
Research Center (QSRC) at Dongguk University and   
by  grant No.R01-2005-000-10352-0 from the Basic Research Program of KOSEF.

\end{document}